# A Novel Strategy for GaN-on-Diamond Device with a High Thermal Boundary Conductance

Fengwen Mu,[1,⊥] Bin Xu,[2,⊥] Xinhua Wang,[1,3,4,⊥] Runhua Gao,[1] Sen Huang,[1,3,4] Ke Wei,[1,] Kai Takeuchi,[5] Xiaojuan Chen,[1] Haibo Yin,[1] Dahai Wang,[1] Jiahan Yu,[6] Tadatomo Suga,[5] Junichiro Shiomi,[2*] Xinyu Liu[1,3,4*]

[1] High-Frequency High-Voltage Device and Integrated Circuits R&D Center, Institute of Microelectronics, Chinese Academy of Sciences, Beijing 100029, China

[2] Department of Mechanical Engineering, The University of Tokyo, Bunkyo, Tokyo 113-8656, Japan

[3] University of Chinese Academy of Sciences, Beijing 100049, China

[4] Key Laboratory of Microelectronic Devices & Integrated Technology, Institute of Microelectronics, Chinese Academy of Sciences, Beijing 100029, China

[5] Collaborative Research Center, Meisei University, Hino-shi, Tokyo 191-8506, Japan

[6] Integrated Circuit Advanced Process R&D Center, Institute of Microelectronics of Chinese Academy of Sciences, Beijing 100029, China

(Dated: July 22, 2021)

To achieve high device performance and high reliability for the gallium nitride (GaN)-based high electron mobility transistors (HEMTs), efficient heat dissipation is important but remains challenging. Enormous efforts have been made to transfer a GaN device layer onto a diamond substrate with a high thermal conductivity by bonding. In this work, two GaN-diamond bonded composites are prepared via modified surface activated bonding (SAB) at room temperature with silicon interlayers of different thicknesses (15 nm and 22 nm). Before and after post annealing process at 800 oC, thermal boundary conductance (TBC) across the bonded interface including the interlayer and the stress of GaN layer are investigated by time-domain thermoreflectance and Raman spectroscopy, respectively. After bonding, the 15 nm Si interlayer achieved a higher TBC. The post-annealing significantly increased the TBC of both interfaces, while the TBC of 22 nm silicon interlayer increased greater and became higher than that of 15 nm. Detailed investigation of the microstructure and composition of the interfaces were carried out to understand the difference in interfacial thermal conduction. The obtained stress was no more than 230 MPa for both before and after the annealing, and this high thermal stability of the bonded composites indicates that the room temperature bonding can realize a GaN-on-diamond template suitable for further epitaxial growth or device process. This work brings a novel strategy of SAB followed by high-temperature annealing to fabricate a GaN-on-diamond device with a high TBC.

## 1 Introduction

GaN-based high electron mobility transistors (HEMTs) can be operated with high voltage, high frequency, and high power density.[1] A further increase of power-density of GaN-based HEMTs is strongly desired in future communications applications, especially as the system miniaturization progresses.[2-3] How to realize an efficient heat-dissipation to weaken the impact of near-junction self-heating is always a challenge for the high-power density GaN-based HEMTs to keep high device reliability and long lifetime.[4-5]

GaN-on-SiC device is the main-stream technique, owing to the high thermal conductivity of SiC and the feasibility of high-quality hetero-epitaxial growth of GaN.[6-9] However, the thermal conductivity of SiC is still not high enough to fully take advantages of the potential of GaN-based devices.[10-15] More attention has been attracted by GaN-on-Diamond devices in recent years, since diamond has a much higher thermal conductivity than SiC.[12-18] In addition, much effort has been spent on the reduction of the thermal boundary resistance of GaN-diamond interfaces to further improve the cooling efficiency of the devices.[13-14, 16-18]

Currently, there are two methods to realize the GaN-diamond integration: growth and bonding.[4-5, 11-12, 15-17] Growth method, which has been extensively studied, mainly includes diamond growth on GaN front-surface and diamond growth on the backside of GaN after the removal of original substrate.[4-5, 15, 17-19] Growth methods have the problem of thermal stress, thermal damage, and nucleation layer with low thermal conductivity, although significant progress has been achieved.[17, 19-22] As an alternative to the growth method as well as its advantage in allowing device-first process, bonding method has also been paid much attention.[11-12, 16, 23-30] D. Francis et al. realized GaN-diamond hydrophilic bonding at a temperature higher than 700 °C using a ~22-nm-thick SiN layer, which brings a thermal boundary conductance (TBC) of ~59 MW/m$^2$K.[23, 27] To reduce the mismatch of thermal expansion coefficient during heterogeneous integration, P. C. Chao et al. demonstrated a low-temperature hydrophilic bonding of GaN-diamond at less than 150 °C.[24] The thickness of the interfacial layer was ~35 nm and TBC was ~29 MW/m$^2$K.[28] T. Liu et al. developed a similar low-temperature GaN-diamond hydrophilic bonding using a 30-40 nm interlayer at 180 °C. However, the device self-heating is still obvious due to a low interfacial TBC.[25] T. Gerrer et al. reported a GaN-diamond integration by contacting them in water followed by annealing at ~200 °C in vacuum. A ~30 nm interlayer mainly consisting of Al and O formed during the bonding at the interface.[29]

Recently, we have reported the room temperature GaN-diamond bonding via modified surface activated bonding (SAB) with Si interlayer.[26] The TBC of room temperature bonded interface has been confirmed to be as high as that formed by a high-temperature growth process.[30] However, the thermal stability of the room temperature GaN-diamond composites has not been investigated. If the interface can withstand a high-temperature treatment, a GaN-on-diamond template suitable for further epitaxial growth or device process can be fabricated by room-temperature bonding and transfer process, which will bring much more impact on the real application. Besides, the influence of interlayer thickness on the TBC and the stress of the GaN layer have not been investigated, especially for samples after a high-temperature process.

In this work, we prepared two kinds of GaN-Diamond samples by room-temperature bonding with interlayers with different thicknesses for comparison. The bonded interfaces before and after annealing were tested by the time-domain thermorefelectance (TDTR) method to evaluate their TBC, and were also characterized to understand the effects of their microstructure and composition on their thermal properties. In addition, Raman spectroscopy was utilized to inspect the stress of the GaN layer on the diamond substrate.

## 2 Material and Methods

Single crystalline CVD diamond with a size of 10 mm×10 mm and a thickness of 500-μm was bonded to templated GaN films at room temperature by modified SAB with Si interlayer deposited by sputtering. Si interlayer with two kinds of thickness (~15 nm and ~22 nm) was selected for comparison. The GaN specimens were ~2-μm-thick Ga-face GaN film on sapphire substrates (~430 μm thick) grown by MOCVD method. All GaN and diamond surfaces were polished till their root-mean-square (RMS) surface roughness was less than 0.5 nm. Then both GaN and diamond surfaces were deposited with Si interlayer, activated by the ar ion beam with a power of 1. 2 kV and 400 mA, and bonded at room temperature. More details of the bonding process have been described in the previous publication.[26] After bonding, the sapphire was lift off by a local laser heating process. An annealing treatment on the two bonded samples was conducted to study the interface thermal stability. The annealing was carried out at 800 °C holding for 15 min in a $N_2$ purged chamber. Both the heating and cooling rate are 1 °C/min.

A typical TDTR method was employed to investigate the thermal boundary conductance (TBC) between the GaN and diamond ($G_{GaN-diamond}$). The GaN surface was further smoothed by chemical-mechanical polishing to obtain a proper TDTR sensitivity. An aluminum (Al) layer with a thickness of 100 nm was deposited on the surface of the GaN as the transducer layer for TDTR. More details of the setup and the operating principle of the TDTR have been described in the previous paper.[37] A femtosecond pulse laser (Ti: Sapphire, pulse width: 140 ps) with a frequency of 80 MHz and a wavelength of 800 nm was used as the source of pump and probe laser. The laser source was split into two ways for pump and probe. The pump laser was first conversed to 400 nm via a BIBO crystal and then modulated with the frequencies of 1.111 MHz or 11.05 MHz before heating the sample surface. A delay stage was employed to adjust the light-path of the probe laser, which can vary the delay time between the pump and probe laser. The delay time of 0-0.7 ns of the $V_{out}$ signal comprises the acoustic echo and was used to identify the thickness of the GaN layer. The –$V_{in}/V_{out}$ signal with delay time ranging from 0.2 ns to 8 ns was used to measure the desired parameters by fitting with the theoretical model.[38,39]

Cross-section TEM samples were prepared with a Thermo Scientific Helios dual beam focused ion beam (FIB) system. HR-STEM (Thermo Scientific Tecnai F20) operated under 200 kV acceleration voltage was used to characterize the interfacial structure. The EDS mapping was collected by a Super-XTM Super Energy Spectrum with a symmetric four-probe and the collection time per pixel was 20 μs. The Raman spectroscopy measurement (Renishaw inVia) was performed using a 532 nm laser and a grating of 3000 to confirm a wavenumber accuracy of 0.02 cm$^{-1}$. The laser power was set down to 1.91 mW, which can exclude the heat-induced shift of Raman peaks (This was further verified by gradually changing the laser power from 1.91 to 30 mW, where a negligible variation of the Raman peaks position was observed).

## 3 Results and Discussion

GaN-on-diamond composite substrate was achieved by bonding a GaN-on-sapphire template to a commercial single-crystalline diamond substrate using two kinds of Si interlayer with thickness of ~15 nm (sample #1) and ~22 nm (sample #2), followed by the removal of the sapphire substrate. The modified SAB with Si layer sputtering-deposition was utilized to bond GaN and diamond. More details about the bonding process are described in the experimental part. An annealing process at ~800 °C is applied on these bonded GaN-diamond composites to simulate an epitaxial growth or ohmic-contact formation process. Consequently, both samples were found to withstand the annealing process.

The TBC of the GaN-diamond interfaces ($G_{GaN-dia}$) was measured by the TDTR method. The aluminum (Al) layer deposited on the GaN-diamond bonded composite, shown in Figure 1 (a), was used as a transducer. As for the parameters in the physical model fitted to the TDTR signal, the thermal conductivity of diamond ($k_{dia}$) is obtained by measuring an Al/diamond reference sample, and the thickness of the transducer layer (Al) is measured using an Al/quartz reference sample, where the Al layer was deposited concurrently with the Al/GaN/diamond samples. The thermal conductivity of Al was referred to the previous reference,[31] since its sensitivity is almost negligible. As for the specific heat of each layer of the thin film, the values of corresponding bulk materials were used because their variations with size are negligible.

Since the modulation frequency is critical in determining the detection depth, measurement was initially carried out using laser with a higher modulation frequency of 11.05 MHz to obtain thermal conductivity of GaN ($k_{GaN}$) and TBC of Al/GaN interface ($G_{Al-GaN}$), followed by the measurement of $G_{GaN-dia}$ with a lower modulation frequency of 1.111 MHz. Both $k_{GaN}$ and $G_{Al-GaN}$ have a considerable sensitivity (see Figure S1) under two corresponding modulation frequencies, thus a highly reliable analysis can be guaranteed. Multiple points were measured for each sample to check the uniformity of the bonding interface. Concerning the unintentionally induced thickness nonuniformity of GaN during the thinning and polishing procedure, we used the acoustic echo signal (0~0.7 ns) to obtain the actual thickness of the GaN layer at each point (Figure S2), which contributes to a proper fitting of the temperature decay profile with the theoretical model.

As a result of measurement under 11.05 MHz, the $k_{GaN}$ of sample #1 approximates 210 W/m-K, higher than that of sample #2 (149.6 W/m-K). As the thickness of GaN of sample #1 (1928~1932 nm) is larger than that of sample #2 (1547~1571 nm), the relatively small $k_{GaN}$ of sample #2 is considered to result from the extra phonon boundary scattering caused by the smaller thickness. This explanation is supported by a previous study suggesting the mean free path of phonon in GaN that predominates the heat conduction ranges from 100 nm to 3 μm.[32] We then conducted the TDTR measurement with 1.111 MHz modulation frequency to measure $G_{GaN-dia}$. The same TDTR measurement using two frequencies was also carried out for the annealed samples. The experimental data and the analytical heat transfer solution matches well, as shown in Figure 2. The $G_{GaN-dia}$ for each as-bonded sample shows an error bar of ±5.6% (Sample #1) and ±2.6% (Sample #2), indicating the SAB bonding in the present study is beneficial for achieving a uniform interface. The $G_{GaN-dia}$ was increased from 32.4 MW/m$^2$-K and 28.0 MW/m$^2$-K to 71.3 MW/m$^2$-K and 85.9 MW/m$^2$-K for sample #1 and #2 after annealing, respectively, as shown in Figure 1 (b). Unexpectedly, the $G_{GaN-dia}$ of sample #2 becomes larger than that of sample #1, opposite from the trend of pristine samples.

The microstructure and composition of the interfaces of sample #1 and sample #2 were analyzed by HR-STEM and EDS. Figure 3 shows cross-section HR-STEM images of the two as-bonded interfaces. The thickness of the deposited Si layers on GaN and diamond sides is equal, which is ~7.5 nm in sample #1 and ~11.0 nm in sample #2. Similar amorphous diamond layers of ~3.0 nm, caused by the ion beam bombardment for surface activation, are found at the interfaces, which have been reported in our previous work.[30] Therefore, it is reasonable that sample #1 exhibits a larger $G_{GaN-dia}$ than sample #2, since a thicker amorphous Si interlayer typically brings a higher thermal resistance. It is worth mentioning that the extra phonon scattering caused by size effect[33] or the interfacial phonon mode modification[34] hardly occurs, because the thickness of the Si layer is more than 10 nm, larger than the mean free path of majority phonon in amorphous Si.[35]

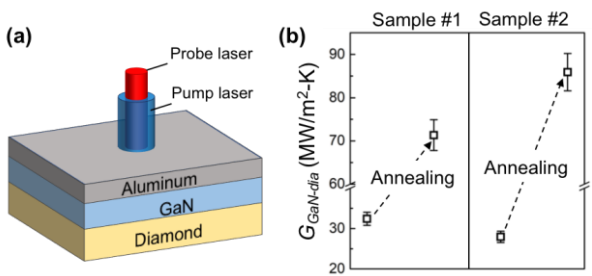

Figure 1. (a) The schematic of TDTR measurement on Al/GaN/diamond structure, (b) TDTR measured thermal boundary conductance of GaN/diamond interface ($G_{GaN-dia}$).

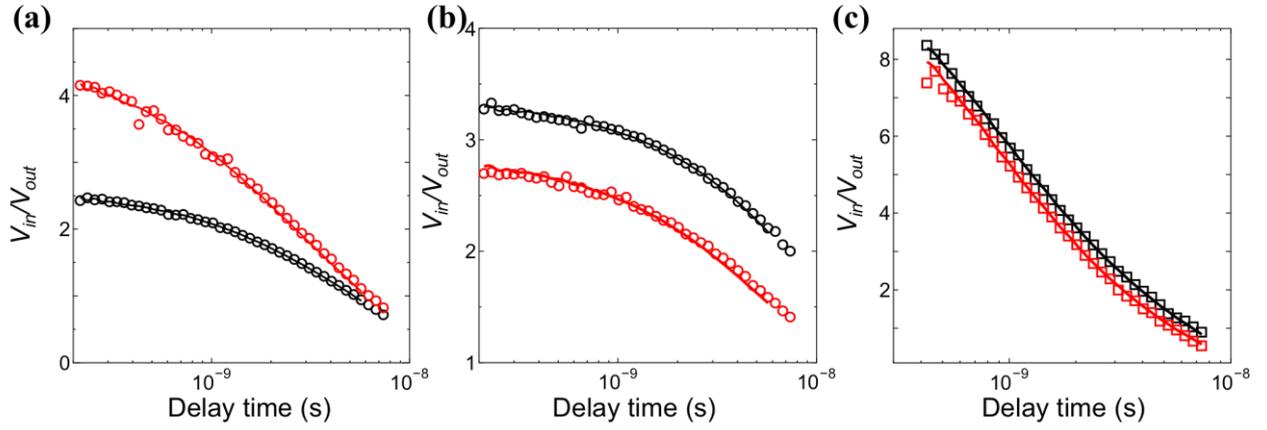

Figure 2. TDTR data fitting of pristine sample #1 and #2 measured with modulation frequency of (a) 11.05 MHz and (b) 1.111 MHz, and annealed samples with modulation frequency of (c) 1.111 MHz. The red markers and curves represent the experimental data and fitting data for sample #2, respectively; the black ones represent that of sample #1.

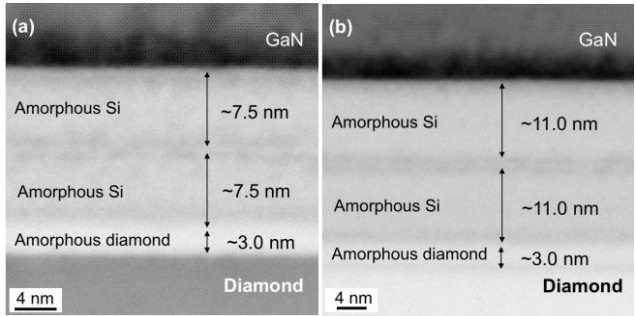

Figure 3. Cross-sectional HR-STEM images of two as-bonded interfaces: (a) sample #1 and (b) sample #2.

According to the EDS results shown in Figure 4, two interfaces exhibit a similar multi-layer element distribution and different thicknesses of the Si layer as shown in the Si map. The Ar element was found at the Si layer/diamond interface, which came from the implantation of the Ar ion beam during surface activation, while it was absent in the GaN side; this phenomenon has been discussed in our previous work.[26] The Fe element originated from the ion beam source made of stainless steel. The O element came from oxidation during bonding and sample preparation for TEM.

Figure 5 illustrates the STEM and EDS results of the annealed samples. After the post annealing, the interlayer of sample #1 was ~18 nm thick, and that of sample #2 was ~27 nm thick. According to the higher-magnification STEM images, as shown in Figure S3, the interfacial amorphous diamond barely crystallized, while the Si interlayer in both samples obviously crystallized. The thickness of the interlayer in sample #1 before and after annealing barely changed, while that in sample #2 thicken after annealing. This thickening is suspected to be caused by the slight oxidation during annealing. By comparing the EDS element-mappings of the interfaces before and after annealing, it is found that the diffusion of Fe, Ar and Ga explicitly occurred and the Fe atoms form clusters at both interfaces. The TBC increase in both samples is assumed to originate from the crystallization of the amorphous Si interlayers and the elimination of the multi-interfaces (Si-Si bonding interface, Si-amorphous diamond interface, etc.) caused by interfacial diffusion. Since the bonding process is exactly the same for both samples, the amount of the introduced Ar and Fe at the two interfaces are considered to be identical; therefore, the concentration of Ar and Fe in the interfacial Si layer of sample #1 is higher than that of sample #2. Besides, the concentration of the diffused Ga in the interfacial Si layer of sample #1 is higher than that of sample #2. Thus, the interfacial disorder extent of the annealed sample #1 is considered higher than that of sample #2, contributing to the less TBC increase compared to sample #2.[36]

Moreover, since the lifetime and the performance of the GaN device are closely correlated with internal stress, it is of significant importance to check the stress possibly induced by the annealing treatment. Raman spectroscopy was applied to evaluate the stress variation inside the GaN layer before and after annealing. The stress in the c-plane of GaN can be evaluated by the shift of Raman peak approximately at 569.6 cm$^{-1}$, which correlates to the atomic oscillation in the c-plane. Typically, 1 GPa of stress would induce a peak shift of 2.56 cm$^{-1}$, where the blueshift and redshift represent the tensile and compression stress, respectively. We performed the Raman mapping on the top surface of GaN, as shown in Figure 6. The mapping result indicates a relatively uniform distribution of compression stress for pristine sample #1 and #2, where the stress is both ~23.6 MPa. After the annealing treatment, the compression stress change to tensile stress, whose value is 141.2 MPa (sample #1) and 221.4 MPa (sample #2). The introduction of tensile stress by the annealing process is caused by the mismatch of the thermal expansion coefficient between diamond and GaN. Notably, according to the Raman spectroscopy measurement, a minor but nonnegligible difference in the Si peaks between these two samples is observed (shown in Figure S4), where the peak of sample #1 is slightly broader than that of sample #2. This broadening of the Si peak of sample #1 may be caused by the larger stress, the smaller gain size, or the more amorphous phase in the Si interlayer, which would also bring a low TBC. However, the cause of the TBC difference may be far more complicated considering the complexity of the interfacial structure and the constituent elements.

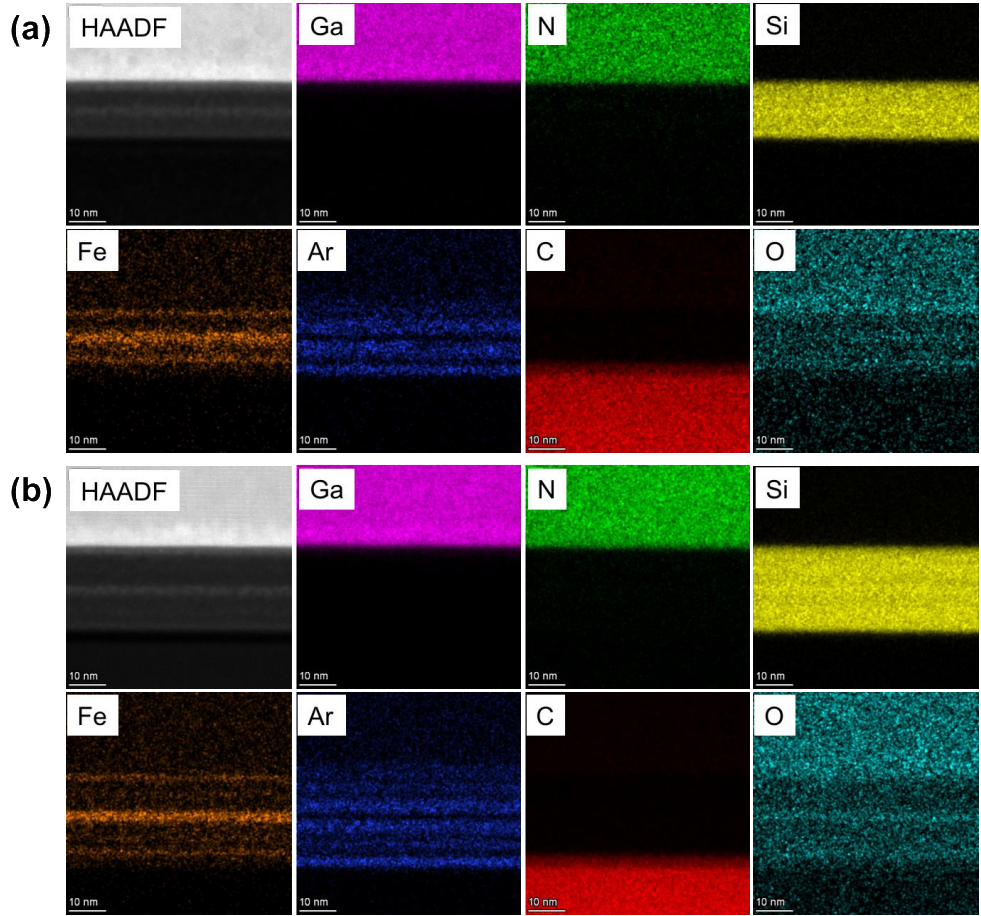

Figure 4. High-resolution high-angle annular dark-field (HAADF) STEM images and EDS element mapping of two as-bonded interfaces:(a) sample #1 and (b) sample #2. HAADF image in gray, Ga map in magenta, N map in green, Si map in yellow, Fe map in orange, Ar map in blue, C map in red, and O in cyan.

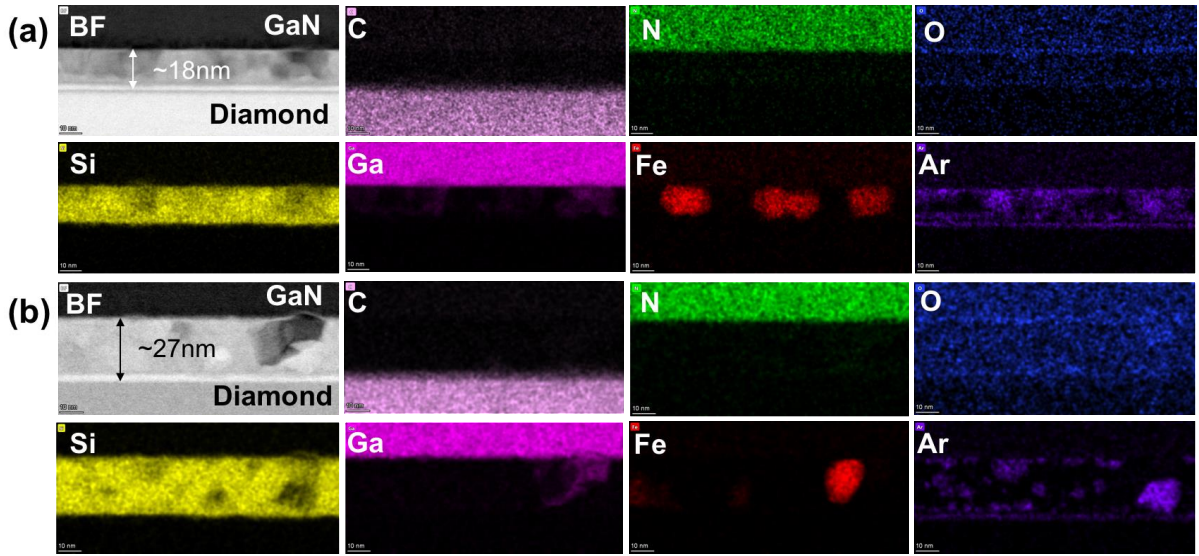

Figure 5. Cross-sectional STEM Bright-field (BF) images and EDS mapping of two interfaces after annealing: (a) sample #1 and (b) sample #2. BF image in gray, C map in pink, N map in green, O in blue, Si map in yellow, Ga map in magenta, Fe map in red, and Ar map in purple.

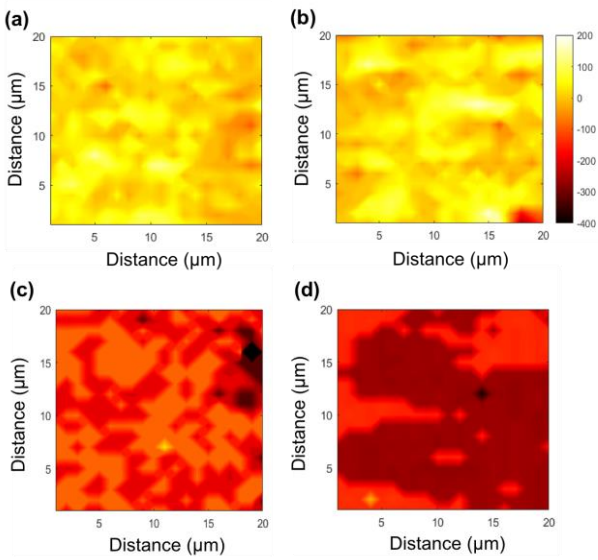

Figure 6. Stress mapping (20 μm×20 μm) of GaN by Raman spectroscopy for pristine (a) sample #1, (b) sample #2, and (c) sample #1, (d) sample #2 with annealing treatment.

Although the stress of the GaN layer is mostly owing to the difference of thermal expansion coefficient $\Delta\alpha$ between diamond and GaN (~4.59×10$^{-6}$ W/K), the resultant strain of the GaN layer (2.99×10$^{-4}$ and 4.68×10$^{-4}$ for sample #1, and #2, respectively, calculated from the stress data) is one order of magnitude smaller than that estimated by $\varepsilon = \Delta\alpha\Delta T$ (~3.6×10$^{-3}$), where the $\Delta T$ is the difference between the annealing temperature and room temperature (800 °C). Therefore, we assume that the room temperature bonding process followed by a post-annealing process (likely to be the electrode formation process) has a remarkable advantage compared to the direct epitaxial growth method. Besides, the Si interlayer may work as a buffer layer to release the stress caused by the mismatch of thermal expansion coefficient.

## 4 Conclusions

In summary, we found the GaN-Diamond bonded composites via modified SAB with ~15 nm or ~22 nm Si interlayer at room temperature can withstand a post-annealing process of 800 °C. The advantage of a ~15 nm Si interlayer to achieve a high TBC for room temperature bonded GaN-diamond interface is confirmed. Annealing at 800 °C increased the TBC of the interfaces with both ~15 nm and ~22 nm Si layers owing to the crystallization of Si-layer and the elimination of the multi-interfaces caused by the interfacial diffusion. Meanwhile, the TBC increase of the interface with a ~15 nm Si layer is much less than that of the interface with a ~22 nm Si layer, leading to that the interface of ~22 nm Si layer has a higher TBC. This difference is assumed to be caused by a synergy effect of the different concentration of Ar, Fe and Ga, the crystallinity difference, and the stress condition in the Si interlayer. Besides, the room temperature bonding introduces stress less than 30 MPa, and the post-annealing introduces stress no more than 230 MPa. It is concluded that the room-temperature bonding followed by a high-temperature process exhibits a remarkable advantage in terms of stress, compared to the conventional high-temperature epitaxial growth method. This work sheds light on a novel strategy for the fabrication of GaN-on-diamond devices with a high interfacial TBC. Furthermore, N-face GaN-on-diamond devices become technically feasible via a structure with a designed epitaxial layer before bonding.

**Acknowledgements**

F. M., X. W., R. G., S. H., K. W., J. Y. and X. L. would like to acknowledge the financial support from National Natural Science Foundation of China (No. 62004213, No. 61534007, 61527816, 61822407, 62074161, 11634002, and 61631021), from the Key Research Program of Frontier Sciences, Chinese Academy of Sciences (CAS) (No. QYZDB-SSW-JSC012), from the National Key R&D Program of China (No. 2016YFB0400105, 2017YFB0403000), from the Youth Innovation Promotion Association of CAS, from the University of Chinese Academy of Sciences, and from the Opening Project of Key Laboratory of Microelectronic Devices & Integrated Technology, Institute of Microelectronics, CAS.
B. X. and J. S. would like to acknowledge the financial support from JSPS KAKENHI (19H00744) and JST CREST (JPMJCR20Q3, JPMJCR19I2).

**Author Contribution Statement**

⊥ These authors contribute equally